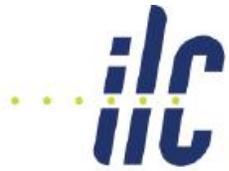
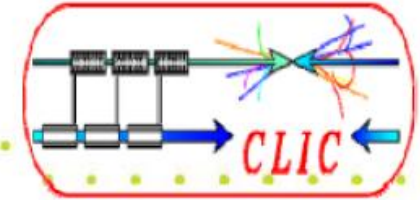

# ASSESSING RISK IN COSTING HIGH-ENERGY ACCELERATORS: FROM EXISTING PROJECTS TO THE FUTURE LINEAR COLLIDER

Ph. Lebrun, CERN, Geneva, Switzerland
P.H. Garbincius, Fermilab, Batavia, IL 60510 USA

*Abstract*

High-energy accelerators are large projects funded by public money, developed over the years and constructed via major industrial contracts both in advanced technology and in more conventional domains such as civil engineering and infrastructure, for which they often constitute one-of markets. Assessing their cost, as well as the risk and uncertainty associated with this assessment is therefore an essential part of project preparation and a justified requirement by the funding agencies. Stemming from the experience with large circular colliders at CERN, LEP and LHC, as well as with the Main Injector, the Tevatron Collider Experiments and Accelerator Upgrades, and the NOvA Experiment at Fermilab, we discuss sources of cost variance and derive cost risk assessment methods applicable to the future linear collider, through its two technical approaches for ILC and CLIC. We also address disparities in cost risk assessment imposed by regional differences in regulations, procedures and practices.

# ASSESSING RISK IN COSTING HIGH-ENERGY ACCELERATORS: FROM EXISTING PROJECTS TO THE FUTURE LINEAR COLLIDER

Ph. Lebrun, CERN, Geneva, Switzerland
P.H. Garbincius, Fermilab, Batavia, IL 60510, U.S.A.

*Abstract*

High-energy accelerators are large projects funded by public money, developed over the years and constructed via major industrial contracts both in advanced technology and in more conventional domains such as civil engineering and infrastructure, for which they often constitute one-of markets. Assessing their cost, as well as the risk and uncertainty associated with this assessment is therefore an essential part of project preparation and a justified requirement by the funding agencies. Stemming from the experience with large circular colliders at CERN, LEP and LHC, as well as with the Main Injector, the Tevatron Collider Experiments and Accelerator Upgrades, and the NOvA Experiment at Fermilab, we discuss sources of cost variance and derive cost risk assessment methods applicable to the future linear collider, through its two technical approaches for ILC and CLIC. We also address disparities in cost risk assessment imposed by regional differences in regulations, procedures and practices.

## INTRODUCTION

Over their century of existence, high-energy particle accelerators have undergone sustained development of their performance, as exemplified in the Livingston diagram [1]. The corresponding increases in size and cost however proceeded at slower pace thanks to implementation of novel technologies and application of industrial construction methods. Still, recent large accelerator projects such as LEP and LHC have costs amounting to several years of funding of the discipline, therefore drawing significantly on public research budgets in their construction years. Assessing risk in costing such projects is therefore an important issue, whether these risks are later mitigated by project de-scoping, stretching of construction schedules or reallocation of additional resources. The assessment is rendered more difficult by the fact that these projects are usually one-of or single-time activity, without a market outside the project proper enabling the establishment of real market prices. Fortunately for the costing engineer, high-technology accelerator components account only for a fraction of the total cost. Much of the budget goes into civil engineering, infrastructure and services for which market prices are usually available (Table 1).

We first discuss the diverse cost variance factors and the ways to handle them, illustrated by examples taken from recent projects. Considering conditions imposed by funding agencies and governing bodies, we then present the implementation of cost risk assessment for the two approaches to the linear collider, i.e. ILC and CLIC.

Table 1: Cost structure of LEP and LHC

| Project | LEP | LHC |
|---|---|---|
| Accelerator components | 30 % | 66 % |
| Accelerator infrastructure | 18 % | 8 % |
| Civil engineering | 43 % | 13 % |
| Injectors | 9 % | 11 % |

## COST VARIANCE FACTORS

Cost risk assessment begins with the identification and understanding of the different factors giving rise to cost variance. Our discussion follows the time line from design to industrial procurement.

A first class of factors pertain to the technical definition of the project: the configuration may still be evolving at the time the cost estimate is performed, and the maturity of the design may be incomplete, imposing to base the estimate on prototype development costs. This affects the work breakdown structure, the quantities and the value of unit costs. Moreover, technological breakthroughs and changes of applicable regulations may well appear by the time the components are produced, thus leading to additional uncertainty. Still, all these effects are essentially of technical nature, and controlled or at least known by the project engineer.

A second class of variance factors appears when going for industrial procurement. If this is handled correctly, the technical specification issued to industry should remove all uncertainties pertaining to the technical definition of the procured equipment or service, discussed in the previous class. The remaining cost variance then only stems from the response of industry, involving both technical and commercial aspects. The price effectively paid for the procurement contract depends on the qualification and experience of the vendor, on the completion of development and industrialization efforts, on the ownership of the design and level of guarantee on the finished delivery asked from the supplier (e.g. build-to-print vs. functional specifications), on the structure and state of the market and on the purchasing rules applied (e.g. lowest bidder vs. best value for money). Further away from the control of the project, they also depend on the commercial strategy of the vendor, balancing the benefits of this contract against competing productions which may generate more profit or ensure longer-term markets, or wishing to penetrate a new market or to have its brand name attached to a highly visible, high-technology project.

Outside the control of the project, a third class of cost variance factors comes from the variations of the general economic situation and of national laws and regulations. This impacts the cost of the project through the escalation of raw-material and industrial prices, the fluctuations of currency exchange rates and the differences in applicable labour laws, taxes and custom duties. While the past evolution of these factors is very well documented, their future variations may only be forecast with considerable uncertainty. In any case, they are not specific to the project, and their handling in the cost estimate will be imposed by the governing and funding bodies, and not left to the initiative of the project cost engineer.

## LHC INDUSTRIAL CONTRACTS

The LHC accelerator project, representing some 3.2 billion Swiss francs of industrial contracts, provides a recent source of relevant information for assessing the cost variance stemming from industrial procurement. This procurement was conducted within a set of well-defined purchasing rules, in particular contract adjudication to the bidder having submitted the lowest valid offer.

We have analysed 218 valid offers received from qualified suppliers in response to 48 invitations to tender for accelerator components, based on detailed technical specifications in the fields of mechanical engineering, electromagnets, vacuum technology, cryogenics and instrumentation. Figure 1 shows the observed distribution of tender prices, normalized to the lowest valid offer. The distribution is far from Gaussian, with the tender prices crowding close to the lowest value, and can be reasonably well fitted by an exponential probability distribution function (pdf) with a threshold of 1 (by definition), a mean $m = 1.46$ and thus a standard deviation $\sigma = 0.46$.

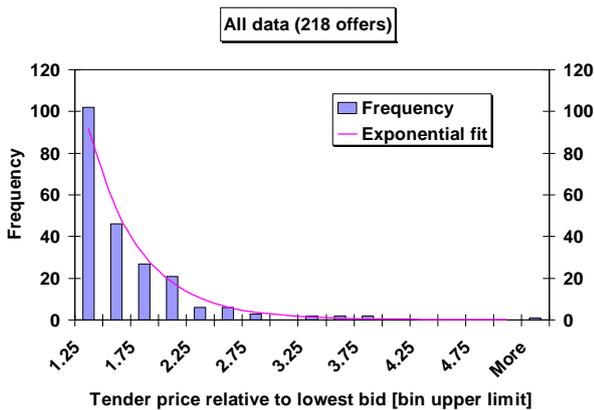

Figure 1: Observed tender prices for LHC accelerator components, normalized to the lowest offer

We postulate this distribution to be representative of the cost variations due to industrial procurement, and we will now use it for estimating the distribution of the lowest offers, i.e. those leading to a procurement contract.

In response to an invitation to tender, consider that $n$ valid offers are received, distributed according to an exponential pdf [$m$, $\sigma$] characterized by the following relation

$$m = 1 + \sigma$$

What is the statistical distribution of the lowest offers among these $n$? Applying the combined-probability theorem and assuming the offers are uncorrelated, yields another exponential distribution of mean $1+\sigma/n$ and standard deviation $\sigma/n$ for the lowest offers among the $n$ received. Thus for LHC accelerator components, with $218/48 \approx 4.54$ valid offers per adjudication on average and a relative standard deviation of 0.46 on tender prices, the relative deviation on contract prices due to industrial procurement can be estimated to $\sigma/n = 0.46/4.54 \approx 0.1$.

## COST RISK ASSESSMENT FOR THE ILC

The Reference Design Report (RDR) for the ILC [2] contains an international VALUE estimate with a $\sigma$ likely to be within the 10-15 % range, and an 95 % confidence level no larger than 25 % above the mean. In addition to the most probable estimate for each cost item, the estimators were asked to present the pdf for the estimate, including the most probable cost, the shape of the pdf - flat, triangular, or (asymmetric) Gaussian - and the parameters of the pdf. The uncertainties or spread of the pdf were often based on the past experience of the estimator. Alternatively, a template (questionnaire) [3], similar to Table 2, was used to estimate the uncertainty based on the maturity of the estimate. This included a weighting based on risks for design, technology, cost estimating methodology, and schedule. The cost uncertainties or risks were simply tabulated to allow each global region to treat their contributions in their local standard manner, whether or not to include a risk budget in the proposal to their particular funding agency.

A guiding principle is that projects worldwide would like to have sufficient financial approval in order to not have to request additional funding from their agencies if some of the risks are realized. Some level of contingency or management reserve is identified and often added to the sum of the most probable estimates to cover up to the 95 % (US DOE) or 98 % (XFEL) [4,5] confidence level for the pdf. The pdf of the cost estimate, and thus the amount added to the estimate to cover this risk budget, depends vitally on the extent of correlations between the uncertainties for the individual cost elements. Ignoring correlations tends to underestimate the risk. Assuming complete 100 % correlations tends to overestimate the risk. The engineers are usually not experienced at estimating the correlations between all cost elements and ensuring that the correlation matrix is self-consistent or non-singular. Lacking specific correlation data, various US agencies (NASA, General Accounting Office, Air Force) recommend using an approximation for a global correlation coefficient of 25-75 %. Although there is no standard guideline for US DOE projects, a linear sum of the $\sigma$ for the cost risks (in currency, not %) is compiled, but is treated as the 95 % confidence level (1.64 $\sigma$). This is equivalent to assuming a 60 % correlation between cost elements for (symmetric) Gaussian uncertainties. Alternatively, finite (between 0 and 100 %) correlations between uncertainties in the cost elements can be estimated, and summed using a self-written Monte-Carlo simulation or a commercial program such as @risk [6].

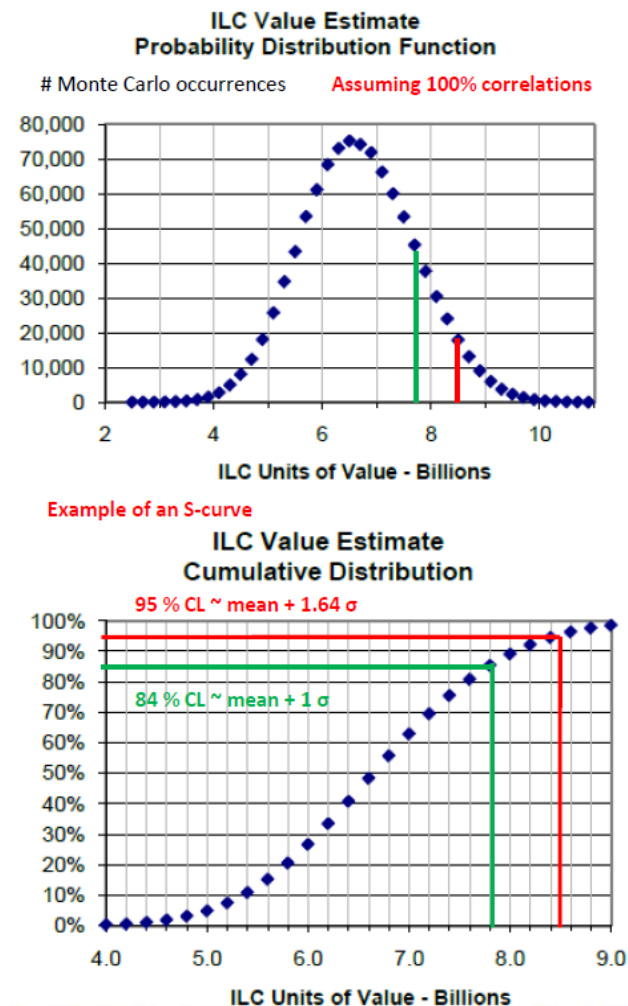

Figure 2: PDF and cumulative distribution for the ILC RDR VALUE estimate, assuming 100% correlations

## COST RISK ASSESSMENT FOR CLIC

The CLIC Conceptual Design Report, presently under work, will include a value cost estimate of the project. The approach used for assessing the uncertainty on the estimate follows from the analysis of cost variance factors presented above. The three classes of factors are assumed statistically independent.

The first class i.e. uncertainty in technical definition, is ranked in three levels defined by the project engineer according to his judgement, each represented by a standard deviation $\sigma_{design}$ conventionally given in Table 2.

The second class i.e. uncertainty in industrial procurement, applies the results of the LHC procurement study. It is conventionally represented by a standard deviation $\sigma_{industry} = 0.5/n$, where $n$ is the number of valid offers expected. $\sigma_{design}$ and $\sigma_{industry}$ are then summed quadratically to obtain the total uncertainty on the cost element.

Table 2: Relative standard deviation due to uncertainty or maturity of the project technical definition

| Technical judgment | $\sigma_{design}$ |
|---|---|
| Known technology | 0.1 |
| Extrapolation from known technology | 0.2 |
| Requires specific R&D | 0.3 |

The third class i.e. escalation and fluctuations of currency exchange rates, is considered outside the scope of project cost assessment. Once the costs are expressed in Swiss francs, the chosen reference currency, they can be escalated in a deterministic way using Swiss industrial indices [7].

In view of the early stage of definition of the CLIC project, it is foreseen to add linearly the absolute uncertainties in the different cost elements, equivalent to assuming full correlation between them, in order to produce a value maximizing the uncertainty on the total cost. Moreover, this uncertainty will only be added to the value cost estimate, so as to provide a measure of contingency.

## ACKNOWLEDGEMENTS

The authors wish to acknowledge the contributions of the ILC Cost Engineers (T. Shidara and W. Bialowons), the ILC Design & Cost Board, F. Lehner, the ILC estimators, G. Riddone and the members of the CLIC Cost & Schedule Working Group to this work, as well as the support of B. Barish and J.P. Delahaye.